\documentclass[12pt,preprint]{aastex}

\begin{document}
\title{CCD photometric study of the contact binary TX Cnc
in the young open cluster NGC 2632 }

\author{Liu L.\altaffilmark{1,2,3} and Qian S.-B.
\altaffilmark{1,2};\\ Soonthornthum, B.\altaffilmark{4}, Zhu,
L.-Y.\altaffilmark{1,2,3},
 He, J.-J.\altaffilmark{1,2,3}, Yuan, J.-Z.\altaffilmark{1,2,3}}

\altaffiltext{1}{National Astronomical Observatories/Yunnan
Observatory, Chinese Academy of Sciences, \\
P.O. Box 110, 650011 Kunming, P.R. China\\
 LiuL@ynao.ac.cn\\
qsb@netease.com}

\altaffiltext{2}{United Laboratory of Optical Astronomy, Chinese
Academy of Sciences (ULOAC),\\ 100012 Beijing, P. R. China}

\altaffiltext{3}{Graduate School of the Chinese Academy of
Sciences,\\ 100012 Beijing, P. R. China }

\altaffiltext{4}{Department of Physics, Faculty of Science, Chiang
Mai University, \\ 50200 Chiang Mai, Thailand \\
boonraks@chiangmai.ac.cn}


\begin{abstract}

TX Cnc is a member of the young open cluster NGC\,2632. In the
present paper, four CCD epochs of light minimum and a complete V
light curve of TX Cnc are presented. A period investigation based on
all available photoelectric or CCD data showed that it is found to
be superimposed on a long-term increase
($dP/dt=+3.97\times{10^{-8}}$\,days/year), and a weak evidence
suggests that it includes a small-amplitude period oscillation
($A_3=0.^{d}0028$; $T_3=26.6\,years$). The light curves in the V
band obtained in 2004 were analyzed with the 2003 version of the W-D
code. It was shown that TX Cnc is an overcontact binary system with
a degree of contact factor $f=24.8\%(\pm0.9\%)$. The absolute
parameters of the system were calculated:
$M_1=1.319\pm0.007M_{\odot}$, $M_2=0.600\pm0.01M_{\odot}$;
$R_1=1.28\pm0.19R_{\odot}$, $R_2=0.91\pm0.13R_{\odot}$. TX Cnc may
be on the TRO-controlled stage of the evolutionary scheme proposed
by Qian (2001a, b; 2003a), and may contains an invisible tertiary
component ($m_3\approx0.097M_{\odot}$). If this is true, the
tertiary component has played an important role in the formation and
evolution of TX Cnc by removing angular momentum from the central
system(Pribulla \& Rucinski, 2006). In this way the contact binary
configuration can be formed in the short life time of a young open
cluster via AML.

\end{abstract}

\section{Introduction}

It is well known that W UMa-type binary stars have a high frequency
in old open cluster (with age no less than 4Gyr). In NGC 188, at
least 9 W UMa-type binaries were reported by Kaluzny \& Shara (1987)
and Zhang et al. (2002, 2004). Four and five W UMa-type stars were
detected in the open clusters M67 and Tombaugh 2
respectively(Gilliland et al. 1991; Sandquist \& Shetrone 2003;
Kubiak et al. 1992), while the two old open clusters, Berkeley 39
and Cr261, were known to possess 12 and at least 28 contact binary
systems, respectively(Kaluzny et al. 1993; Mazur et al. 1995). Mazur
et al. (1995) obtained a lower limit for the frequency of W UMa
binaries in cluster in the range of 1/100-1/60. A high incidence of
W UMa-type binaries correlating with gradually increasing age in old
open clusters is in agreement with the theory of the formation of
contact binary stars via magnetic braking (Rucinski, 1998).
According to this mechanism, a detached system forms a contact
binary by angular momentum loss via magnetic stellar wind, in which
the spin and orbital angular momentum are coupled through tides
(e.g., Huang 1967; Vilhu 1982; Guinan \& Bradstreet 1988). In this
way, contact binary stars are not expected to present in young open
clusters unless there are some other mechanisms inaction.

The W UMa-type binary star, TX Cnc, which is the first contact
binary found in the Praesepe (M44, NGC 2632)cluster, was discovered
to be variable by Haffner (1937) . Complete photoelectric light
curves of the system were derived by Yamasaki \& Kitamura (1972),
Whelan et al. (1973), and Hilditch (1981). Radial velocity curves
and spectroscopic elements were obtained by Popper (1948), Whelan et
al. (1973), McLean \& Hilditch (1983), and Pribulla et al. (2006).
Photometric solutions of TX Cnc were given by several authors (e.g.,
Wilson \& Biermann 1976; Hilditch 1981). It was shown that TX Cnc is
a W-type contact binary system in which the hotter star is the less
massive component. Praesepe is a young open cluster with an age of
$(3-5)\times10^{8}$ years (e.g., Von Hoerner 1957; Maeder 1971).
Bolte (1991) showed that Praesepe contains 10 binary systems, but
only one (TX Cnc) is a contact binary. The presence of TX Cnc in the
young open cluster Praesepe produced interest in it(Guinan \&
Bradstreet 1988; Rucinski 1994), because the fast formation of its
contact configuration is not expected from the theory of angular
momentum loss via magnetic braking. As pointed out by Hazlehurst
(1970),'the occurrence in Praesepe of a W UMa system remains a
paradox'. In this paper, new CCD photometric observations are
presented and the period variations of TX Cnc are analyzed. Then the
triplicity and the evolutionary state of the system are discussed.
We show that TX Cnc may be a triple system with an invisible
companion and thus this paradox can be removed.

\section{New observations for TX Cnc}

TX Cnc was observed on five nights (December 30, 2003; March 16 and
December 18, 19, 2004; March 29, 2006) with the PI1024 TKB CCD
photometric system attached to the 1.0-m reflecting telescope at the
Yunnan Observatory in China. The B and V color systems used are
close to the standard Johnson UBV system. The effective field of
view of the photometric system is $6.5\times6.5$ arc min at the
Cassegrain focus. The integration time for each image before March
2004 is 100\,s, and after that is 50\,s. PHOT (measure magnitudes
for a list of stars) of the aperture photometry package of IRAF was
used to reduce the observed images. The observations obtained on
December 18 \& 19, 2004 are complete in the V band. By calculating
the phase of the observations with Equation 2, the light curves are
plotted (Figure 1) and the original data in the V band are listed in
Table 1. It is shown in this figure that the data are high quality
and the light variation is typical of EW type. Since the light
minimum is symmetric, a parabolic fitting was used to determine the
times of minimum light with a least square method. In all, four
epochs of light minimum were obtained and are listed in Table 2.

\begin{table*}
\begin{minipage}{12cm}
\caption{Photometric Data in the V band for TX Cnc observed with the
1.0 meter telescope at Yunnan observatory}
\begin{tabular}{llllllllll}\hline\hline
JD.Hel.     &${\Delta}m$& JD.Hel.   &${\Delta}m$ &JD.Hel.    &${\Delta}m$&JD.Hel.  &${\Delta}m$& JD.Hel. &${\Delta}m$\\
2453300+    &         &  2453300+   &         &    2453300+  &        &2453300+    &        &  2453300+  &      \\
\hline
   58.1519  &  .068   &   58.2340   & .183    &     58.3215  &  .143  &  58.4047   & .118   &  59.2586   & .245 \\
   58.1539  &  .054   &   58.2363   & .197    &     58.3236  &  .133  &  58.4078   & .134   &  59.2606   & .232 \\
   58.1558  &  .047   &   58.2383   & .201    &     58.3258  &  .116  &  58.4099   & .132   &  59.2626   & .222 \\
   58.1579  &  .042   &   58.2406   & .205    &     58.3278  &  .104  &  58.4120   & .143   &  59.2646   & .208 \\
   58.1599  &  .036   &   58.2427   & .210    &     58.3298  &  .106  &  58.4142   & .151   &  59.2666   & .197 \\
   58.1618  &  .036   &   58.2448   & .223    &     58.3320  &  .095  &  58.4165   & .160   &  59.2686   & .190 \\
   58.1639  &  .036   &   58.2468   & .233    &     58.3340  &  .094  &  58.4188   & .173   &  59.2706   & .179 \\
   58.1659  &  .034   &   58.2489   & .243    &     58.3361  &  .081  &  58.4209   & .179   &  59.2726   & .165 \\
   58.1678  &  .039   &   58.2511   & .260    &     58.3383  &  .075  &  58.4231   & .194   &  59.2747   & .157 \\
   58.1698  &  .030   &   58.2531   & .267    &     58.3403  &  .067  &  58.4250   & .203   &  59.2767   & .147 \\
   58.1718  &  .024   &   58.2551   & .282    &     58.3425  &  .066  &  58.4272   & .212   &  59.2787   & .142 \\
   58.1739  &  .030   &   58.2573   & .292    &     58.3446  &  .060  &  58.4297   & .222   &  59.2808   & .135 \\
   58.1759  &  .029   &   58.2621   & .315    &     58.3468  &  .055  &  58.4320   & .232   &  59.2828   & .124 \\
   58.1780  &  .020   &   58.2641   & .321    &     58.3490  &  .050  &  58.4343   & .245   &  59.2849   & .113 \\
   58.1800  &  .030   &   58.2662   & .324    &     58.3511  &  .052  &  58.4365   & .262   &  59.2870   & .109 \\
   58.1821  &  .027   &   58.2683   & .330    &     58.3533  &  .050  &  58.4385   & .265   &  59.2891   & .105 \\
   58.1841  &  .033   &   58.2703   & .337    &     58.3553  &  .051  &  58.4405   & .286   &  59.2911   & .098 \\
   58.1861  &  .039   &   58.2726   & .337    &     58.3575  &  .045  &  59.2123   & .314   &  59.2931   & .095 \\
   58.1884  &  .043   &   58.2747   & .332    &     58.3596  &  .043  &  59.2143   & .320   &            &      \\
   58.1904  &  .038   &   58.2769   & .332    &     58.3617  &  .041  &  59.2164   & .338   &            &      \\
   58.1923  &  .050   &   58.2789   & .322    &     58.3637  &  .037  &  59.2184   & .344   &            &      \\
   58.1944  &  .051   &   58.2810   & .322    &     58.3658  &  .036  &  59.2205   & .355   &            &      \\
   58.1964  &  .053   &   58.2832   & .309    &     58.3679  &  .037  &  59.2225   & .358   &            &      \\
   58.1985  &  .050   &   58.2852   & .305    &     58.3700  &  .035  &  59.2246   & .361   &            &      \\
   58.2006  &  .053   &   58.2873   & .292    &     58.3720  &  .038  &  59.2265   & .355   &            &      \\
   58.2028  &  .064   &   58.2893   & .284    &     58.3743  &  .041  &  59.2285   & .363   &            &      \\
   58.2047  &  .072   &   58.2914   & .273    &     58.3764  &  .045  &  59.2305   & .362   &            &      \\
   58.2068  &  .074   &   58.2935   & .265    &     58.3782  &  .049  &  59.2324   & .362   &            &      \\
   58.2089  &  .081   &   58.2955   & .253    &     58.3802  &  .044  &  59.2344   & .361   &            &      \\
   58.2109  &  .089   &   58.2975   & .246    &     58.3822  &  .053  &  59.2364   & .349   &            &      \\
   58.2130  &  .098   &   58.2996   & .235    &     58.3843  &  .061  &  59.2384   & .350   &            &      \\
   58.2151  &  .102   &   58.3018   & .222    &     58.3864  &  .070  &  59.2404   & .337   &            &      \\
   58.2172  &  .116   &   58.3040   & .213    &     58.3886  &  .073  &  59.2424   & .337   &            &      \\
   58.2192  &  .119   &   58.3064   & .205    &     58.3905  &  .076  &  59.2444   & .329   &            &      \\
   58.2213  &  .125   &   58.3087   & .197    &     58.3925  &  .083  &  59.2465   & .310   &            &      \\
   58.2235  &  .138   &   58.3109   & .186    &     58.3944  &  .086  &  59.2485   & .302   &            &      \\
   58.2255  &  .148   &   58.3130   & .178    &     58.3965  &  .098  &  59.2506   & .286   &            &      \\
   58.2275  &  .157   &   58.3151   & .165    &     58.3984  &  .097  &  59.2526   & .283   &            &      \\
   58.2296  &  .168   &   58.3172   & .159    &     58.4005  &  .110  &  59.2546   & .271   &            &      \\
   58.2318  &  .174   &   58.3193   & .152    &     58.4025  &  .117  &  59.2566   & .254   &            &      \\
\hline\hline
\end{tabular}
\end{minipage}
\end{table*}

\begin{table*}
\begin{minipage}{12cm}
\caption{New CCD times of light minimum for TX Cnc.}
\begin{tabular}{lllll}\hline\hline
 JD (Hel.)    & Error (days) & Method & Min. & Filters\\
 \hline
2453004.2995 & $\pm0.0003$ & CCD& I    & V      \\
2453081.0629 & $\pm0.0006$ & CCD& II   & V      \\
2453081.0631 & $\pm0.0006$ & CCD& II   & B      \\
2453358.2724 & $\pm0.0002$ & CCD& II   & V      \\
2453358.2725 & $\pm0.0003$ & CCD& II   & B      \\
2453824.0495 & $\pm0.0002$ & CCD& I    & V      \\
\hline\hline
\end{tabular}
\end{minipage}
\end{table*}

\begin{figure}
\begin{center}
\includegraphics[angle=0,scale=1 ]{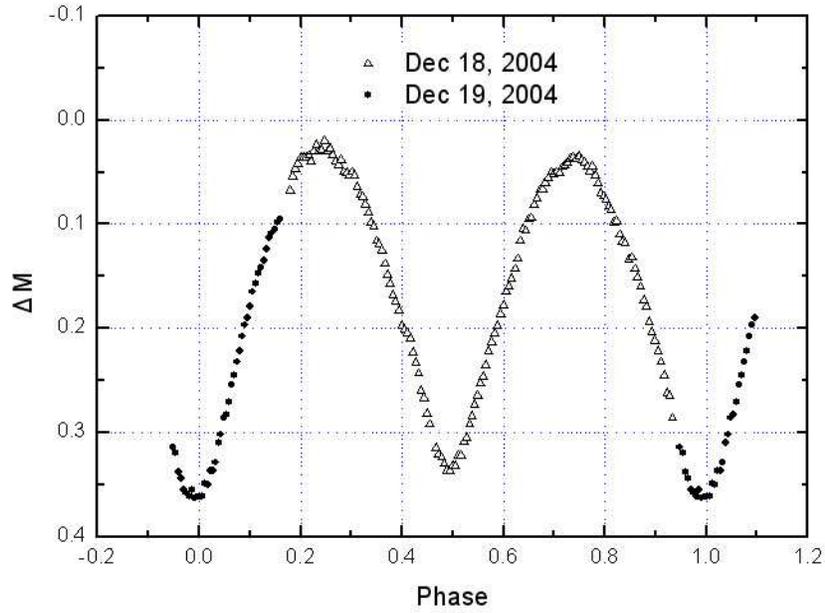}
\caption{CCD data in the V band of TX Cnc observed on 18 and 19
December, 2004.}
\end{center}
\end{figure}

\begin{figure}
\begin{center}
\includegraphics[angle=0,scale=1 ]{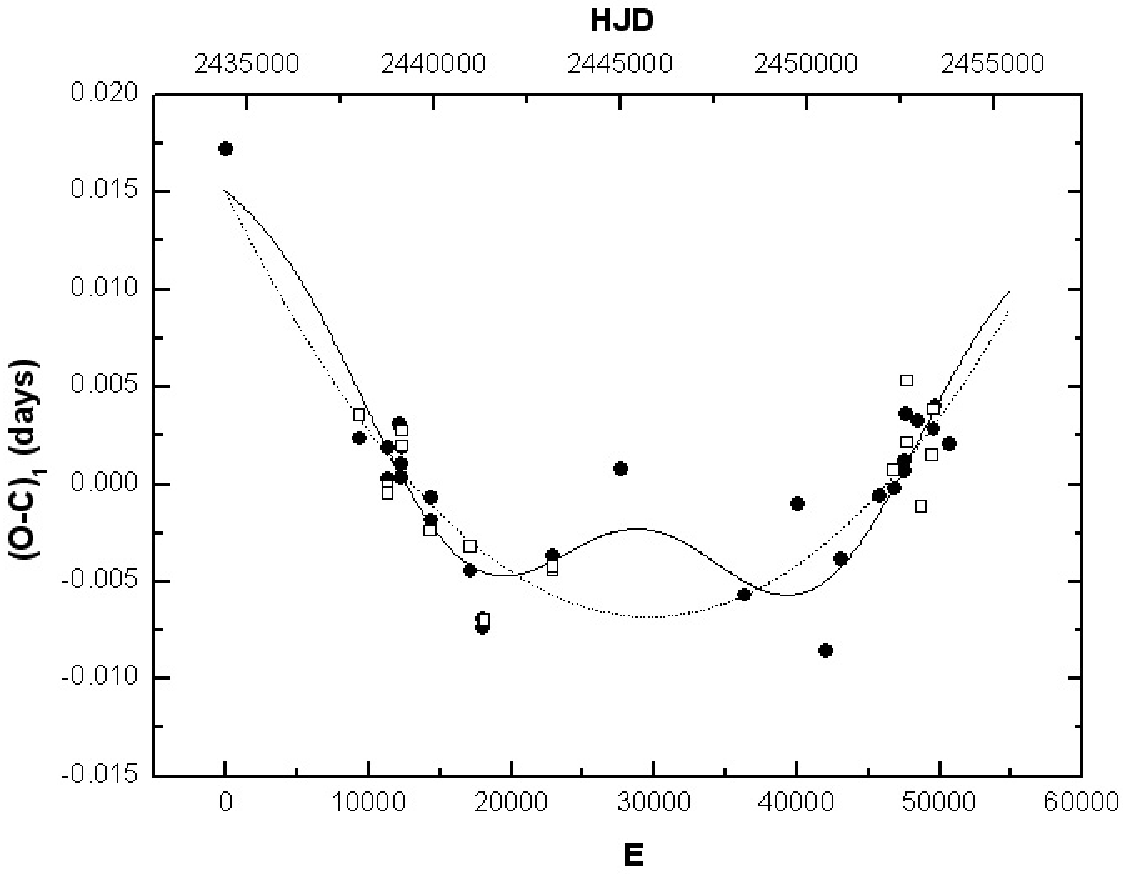}
\caption{ $(O-C)_{1}$ diagram of TX Cnc formed by all available
photoelectric and CCD observations. The $(O-C)_{1}$ values were
computed by using a newly determined linear ephemeris. Solid cycles
refer to the primary minimum and open squares to the secondary
minimum; Solid line represents a combination of a quadratic
ephemeris and a cyclic variation. Also given in dashed line is the
quadratic fit.}\end{center}
\end{figure}

\section{Orbital period variations for TX Cnc}

The orbital period of TX Cnc was first reported to be variable by
Yamasaki \& Kitamura (1972). They collected 20 light minima and
pointed out that a sudden period increase occurred around 1959.
Pribulla et al. (2002) suggested that the period of TX Cnc is
increasing. Qian (2001a) derived a quadratic ephemeris
\begin{eqnarray}
{\rm Min.~I}&=&2434426.4761+0.38288070\times{E}\nonumber\\
    & &+2.94\times{10^{-11}}\times{E^{2}},
\end{eqnarray}

\noindent and a continuous period increase rate of
$dP/dt=+5.61\times{10^{-8}}$\,days/year. In order to investigate the
period change of TX Cnc in detail, all available photoelectric and
CCD observations at times of light minimum were compiled and listed
in Table 3. Based on all collected eclipse times, a new linear
ephemeris was obtained:
\begin{eqnarray}
{\rm Min.~I}&=&2434426.4601(\pm0.0013)\nonumber\\
    & &+0.^{d}38288238(\pm0.00000004)\times{E}.
\end{eqnarray}
\noindent The $(O-C)_{1}$ values with respect to the linear
ephemeris are listed in the fifth column of Table 3. The
corresponding $(O-C)_{1}$ diagram is displayed in Figure 2.

The general $(O-C)_{1}$ trend of TX Cnc, shown in Figure 2,
indicates the continuous period increase reported by Pribulla et al.
(2002) and Qian (2001a). However, a long-term period increase alone
(dashed line in Figure 2) cannot describe the $(O-C)_{1}$ curve very
well, and there is possibly very weak evidence that there may be a
small-amplitude period oscillation. Assuming that the period
oscillation is cyclic, then, based on a least-square method, a
sinusoidal term is added to a quadratic ephemeris to give a better
fit to the $(O-C)_{1}$ curve (solid line in Figure 2). The result is
\begin{eqnarray}
{\rm Min.~I}&=&2434426.4740(\pm0.0001) \nonumber\\
    &  & +0.^{d}38288113(\pm0.00000001)\times{E}\nonumber\\
    & &+2.08(\pm0.22)\times{10^{-11}}\times{E^{2}}\nonumber\\
    & &+0.0028(\pm0.0006)\sin[0.^{\circ}0142\times{E}\nonumber\\
    & &+32.^{\circ}6(\pm0.^{\circ}04)].
\end{eqnarray}

\noindent With the quadratic term in this equation, a secular period
increase rate is determined,
$dP/dt=+3.97\times{10^{-8}}$\,days/year, which is close to the value
derived by Qian (2001a).

The $(O-C)_{2}$ values with respect to the quadratic ephemeris in
Eq.(3) are shown in Figure 3. Although the data after E=27500 show
large scatters, a small-amplitude oscillation is disputably seen in
this figure. However, as we will see, there are a lot of scattered
data points, especially around E=27500. In spite of this, by using
this relation,
\begin{equation}
\omega=360^{\circ}P_{e}/T,
\end{equation}
\noindent where $P_{e}$ is the ephemeris period ($0.^{d}38288238$),
the period of the orbital period oscillation is determined to be
T=26.6\,years. Nevertheless, it is not reliable to rely on just a
few points, so further observations and studies will be needed.
\begin{table*}
\caption{Photoelectric and CCD times of light minimum for TX Cnc.}
\begin{center}
\begin{small}
\begin{tabular}{lllllll}\hline\hline
JD.Hel.  &  Min.  &  Method  &  E  &  $(O-C)_{1}$ & $(O-C)_{2}$ & Ref.$^{*}$\\
2400000+   &    &     &          &             &              &\\
\hline
34426.4773 & I  & pe  &  0       & +0.01719    & +0.00225     & (1)\\
38011.2000 & II & pe  &  9362.5  & +0.00357    & +0.00020     & (2)\\
38012.1560 & I  & pe  &  9365    & +0.00236    & -0.00099     & (2)\\
38774.0915 & I  & pe  &  11355   & +0.00192    & +0.00046     & (2)\\
38774.2809 & II & pe  &  11355.5 & -0.00011    & -0.00156     & (2)\\
38775.0463 & II & pe  &  11357.5 & -0.00048    & -0.00193     & (2)\\
38775.2385 & I  & pe  &  11358   & +0.00027    & -0.00118     & (2)\\
39095.3310 & I  & pe  &  12194   & +0.00310    & +0.00238     & (2)\\
39141.0843 & II & pe  &  12313.5 & +0.00196    & +0.00134     & (2)\\
39141.2748 & I  & pe  &  12314   & +0.00101    & +0.00039     & (2)\\
39142.9995 & II & pe  &  12318.5 & +0.00274    & +0.00212     & (2)\\
39143.1885 & I  & pe  &  12319   & +0.00030    & -0.00031     & (2)\\
39153.1450 & I  & pe  &  12345   & +0.00186    & +0.00127     & (2)\\
39920.0547 & I  & pe  &  14348   & -0.00184    & -0.00081     & (2)\\
39921.9703 & I  & pe  &  14353   & -0.00066    & +0.00037     & (2)\\
39922.1600 & II & pe  &  14353.5 & -0.00240    & -0.00136     & (2)\\
40986.9551 & II & pe  &  17134.5 & -0.00321    & -0.00026     & (2)\\
40987.1453 & I  & pe  &  17135   & -0.00445    & -0.00150     & (2)\\
41331.7365 & I  & pe  &  18035   & -0.00739    & -0.00390     & (3)\\
41332.8856 & I  & pe  &  18038   & -0.00694    & -0.00345     & (3)\\
41372.1310 & II & pe  &  18140.5 & -0.00698    & -0.00343     & (2)\\
43191.7828 & I  & pe  &  22893   & -0.00371    & +0.00197     & (4)\\
43192.7393 & II & pe  &  22895.5 & -0.00442    & +0.00127     & (4)\\
43200.7800 & II & pe  &  22916.5 & -0.00425    & +0.00144     & (4)\\
45022.3480 & I  & pe  &  27674   & +0.00080    & +0.00752     & (5)\\
48332.3597 & I  & pe  &  36319   & -0.00570    & +0.00001     & (6)\\
49777.3625 & I  & pe  &  40093   & -0.00101    & +0.00309     & (7)\\
50515.5522 & I  & CCD &  42021   & -0.00855    & -0.00553     & (8)\\
50926.3897 & I  & CCD &  43094   & -0.00385    & -0.00151     & (9)\\
51952.5177 & I  & CCD &  45774   & -0.00063    & -0.00026     & (10)\\
52348.4185 & I  & pe  &  46808   & -0.00022    & -0.00070     & (11)\\
52352.4397 & II & pe  &  46818.5 & +0.00071    & +0.00021     & (11)\\
52611.8430 & I  & CCD &  47496   & +0.00119    & +0.00010     & (12)\\
52647.8334 & I  & CCD &  47590   & +0.00065    & -0.00051     & (13)\\
52685.3588 & I  & CCD &  47688   & +0.00357    & +0.00231     & (14)\\
52691.2952 & II & CCD &  47703.5 & +0.00530    & +0.00403     & (15)\\
52711.9677 & II & CCD &  47757.5 & +0.00215    & +0.00083     & (16)\\
53004.2995 & I  & CCD &  48521   & +0.00325    & +0.00123     & (18)\\
53081.0630 & II & CCD &  48721.5 & -0.00116    & -0.00336     & (18)\\
53358.2725 & II & CCD &  49445.5 & +0.00148    & -0.00141     & (18)\\
53410.5373 & I  & CCD &  49582   & +0.00284    & -0.00019     & (17)\\
53422.9820 & II & CCD &  49614.5 & +0.00386    & +0.00079     & (16)\\
53455.3357 & I  & CCD &  49699   & +0.00400    & +0.00085     & (17)\\
53824.0495 & I  & CCD &  50662   & +0.00206    & -0.00206     & (18)\\
\hline\hline
\end{tabular}
\end{small}
\end{center}
\noindent {\small $^{*}$ (1) Lenouvel \& Daguillon (1956); (2)
Yamasaki \& Kitamura (1972); (3) Whelan, Worden and Mochnacki
(1973); (4) Hilditch (1981); (5) Diethelm (1982); (6) Diethelm
(1991); (7) Diethelm (1995); (8) Krobusek (1997); (9) Diethelm
(1998); (10) Diethelm (2001); (11) Pribulla et al. (2002); (12)
Dvorak (2003); (13) Nelson (2004); (14) Hubscher (2005); (15)
Diethelm (2003); (16) Kim et al. (2006); (17) Hubscher et al.
(2005); (18) The present authors.\\}
\end{table*}

\begin{figure}
\begin{center}
\includegraphics[angle=0,scale=1 ]{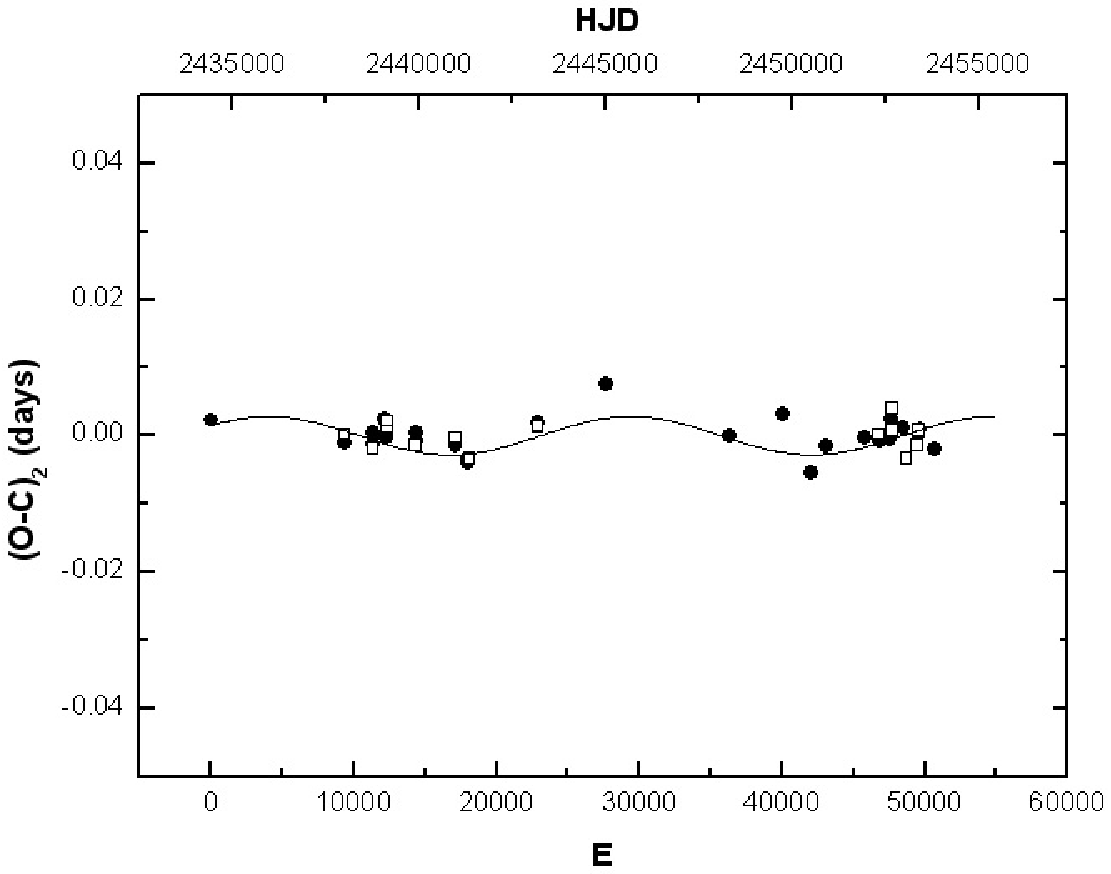}
\caption{$(O-C)_{2}$ values for TX Cnc with respect to the quadratic
ephemeris in Eq.(3).The symbols are the same as figure 2. Solid line
refers to the theoretical orbit of an assumed third body.}
\end{center}
\end{figure}

\section{Photometric Solution}

Photometric parameters of TX Cancri have been derived by several
authors, e.g., Whelan et al. (1973), Wilson \& Biermann (1976) and
Hilditch (1981). All of them found the mass ratio is near 0.6
(Whelan et al. 1973; Wilson \& Biermann, 1976), while Pribulla \&
Rucinski et al. (2006) obtained the spectroscopic mass ratio,
$q_{sp}=0.455\pm0.011$. To check this value, a q-search method with
the 2003 version of the W-D program (Wilson \& Devinney, 1971;
Wilson, 1990, 1994; Wilson \& Van Hamme, 2003) was used (Figure 4).
Firstly, we fixed q to 0.2, 0.3, 0.4 and so on, as figure 4 (the
left one) shows. It can be seen that the best result is obtained
with a value of $q=2$. Secondly, we performed additional solutions
around this value, and found that the best mass ratio is $q=2.3$,
which agrees very well with the results of Pribulla et al. (2006).

During the solution, the mass ratio is fixed on the spectroscopic
value 2.1978 that was obtained by Pribulla et al. (2006). The same
value of temperature for star 1 (star eclipsed at secondary light
minimum) as that used by Wilson \& Biermann (1976) ($T_1=6400$K) was
chosen. The bolometric albedo $A_1=A_2=0.5$ (Rucinski 1969) and the
values of the gravity-darkening coefficient $g_1=g_2=0.32$ (Lucy
1967) were used, which correspond to the common convective envelope
of both components. A limb-darkening coefficient of 0.62 in V was
used, according to Claret \& Gimenez (1990). The adjustable
parameters were: the orbital inclination $i$; the mean temperature
of star 2, $T_2$; the monochromatic luminosity of star 1, $L_{1V}$;
and the dimensionless potential of star 1 ($\Omega_1=\Omega_2$, mode
3 for contact configuration). The O'Connell effect of the system is
so obvious, and as TX Cnc's spectral type is G0-G1V (Yamasaki \&
Kitamura, 1972), F8V (Popper 1948), or F6 (Haffner \& Heckmann,
1937), (a later type), it seemed that chances were good that
starspots will appear on the surface of the star. In that case, we
add a dark spot on the more massive component (the cold one) as many
researchers have done (e.g., Binnedijk, 1960, Mullan, 1975, Bell, et
al., 1990, Linnell \& Olson, 1989). Mullan (1975) deemed that dark
spots exist in contact binaries due to their deep convective
envelopes. The photometric solutions are listed in Table 4 and the
theoretical light curves computed with those photometric elements
are plotted in Figure 5, meanwhile, the geometrical structure of TX
Cnc is displayed in Figure 6.

\begin{figure}
\begin{center}
\includegraphics[angle=0,scale=1 ]{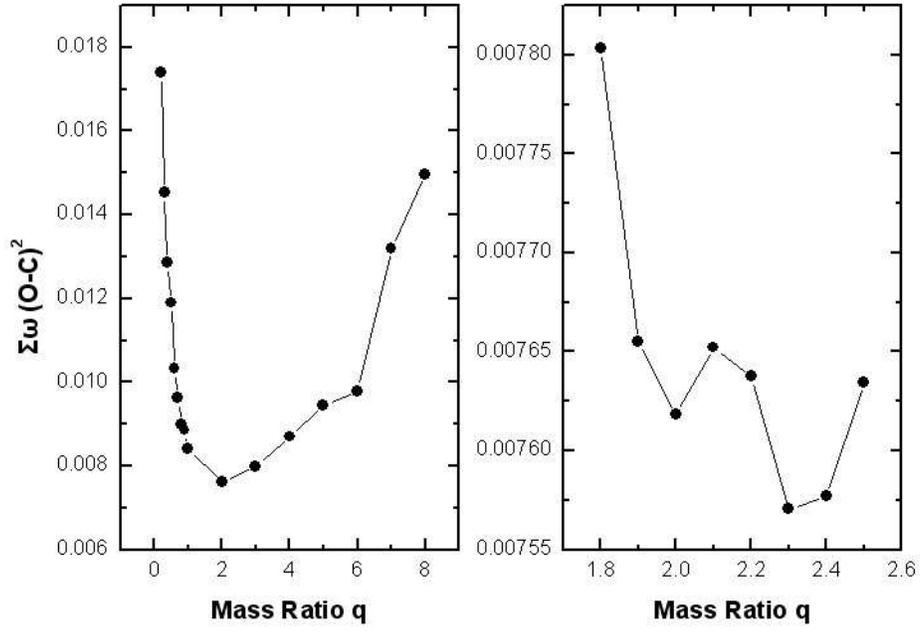}
\caption{The relation between q and $\Sigma$ for TX Cnc. The figure
on the right shows more detail around the best mass ratio $q=2$. }
\end{center}
\end{figure}

\begin{figure}
\begin{center}
\includegraphics[angle=0,scale=1 ]{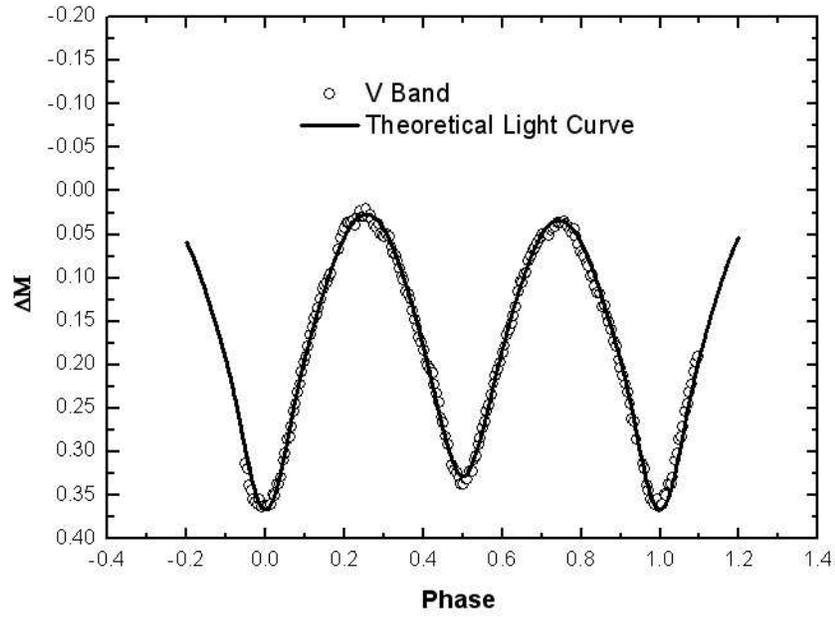}\caption{Observed and
theoretical light curves in the V band for TX Cnc with a spot on the
more massive component.}\end{center}
\end{figure}

\begin{figure}
\begin{center}
\includegraphics[angle=0,scale=1 ]{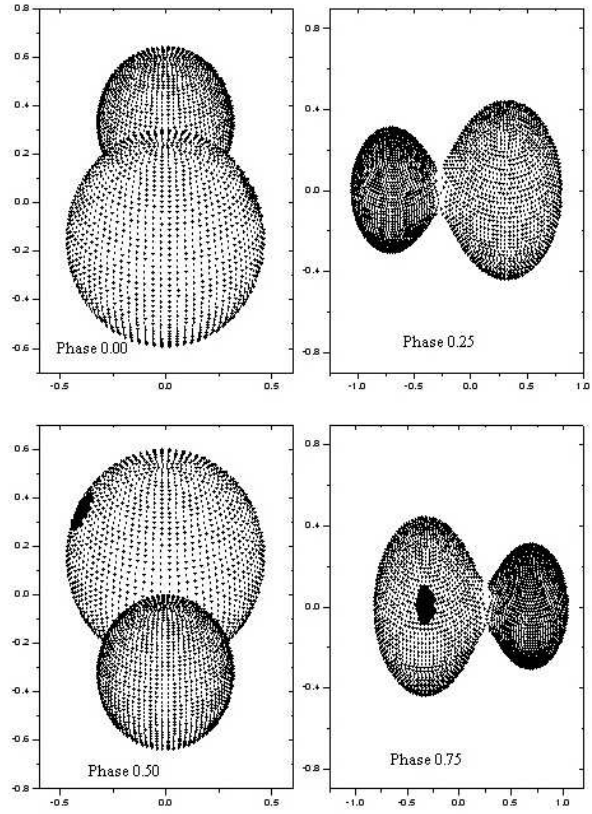}\caption{Geometrical
structure of the open cluster's contact binary TX Cnc with a dark
spot on the more massive component at phase 0.00, 0.25, 0.50 and
0.75.} \end{center}
\end{figure}

\begin{table*}
\caption{Photometric Solutions for TX Cnc.}
\begin{tabular}{lclcl}\hline\hline
Parameters              &  Photometric elements &  errors  & Photometric elements &  errors\\
                        &  with dark spot       &          & without dark spot    &        \\
\hline
$g_1=g_2$               &    0.32               & assumed  &    0.32               & assumed\\
$A_1=A_2$               &    0.50               & assumed  &    0.50               & assumed\\
$x_{1V}=x_{2V}$         &    0.62               & assumed  &    0.62               & assumed\\
$T_1$                   &    6400K              & assumed  &    6400K              & assumed\\
$q$                     &    2.1978             & assumed  &    2.1978             & assumed\\
$\Omega_{in}$           &    5.5292             & --       &    5.5292             & --    \\
$\Omega_{out}$          &    4.9264             & --       &    4.9264             & --     \\
$T_2$                   &    6058K              & $\pm19$K &    6058K              & $\pm21$K \\
$i$                     &    62.241             & $\pm0.31$&    62.015             & $\pm0.18$ \\
$L_1/(L_1+L_2)(V)$      &    0.3853             & $\pm0.0435$ & 0.3876             & $\pm0.0405$ \\
$\Omega_1=\Omega_2$     &    5.3796             & $\pm0.0083$ & 5.3929             & $\pm0.0066$\\
$r_1(pole)$             &    0.3051             & $\pm0.0007$ & 0.3039             & $\pm0.0006$\\
$r_1(side)$             &    0.3202             & $\pm0.0009$ & 0.3188             & $\pm0.0007$\\
$r_1(back)$             &    0.3623             & $\pm0.0015$ & 0.3599             & $\pm0.0012$\\
$r_2(pole)$             &    0.4342             & $\pm0.0007$ & 0.4331             & $\pm0.0005$\\
$r_2(side)$             &    0.4650             & $\pm0.0009$ & 0.4636             & $\pm0.0007$\\
$r_2(back)$             &    0.4975             & $\pm0.0012$ & 0.4956             & $\pm0.0010$\\
$f$                     &    $24.8\,\%$         & $\pm0.9\,\%$& $22.6\,\%$         & $\pm0.8\,\%$\\
$\theta$ ($^\circ$)     &    28.86              &             &   --                             \\
$\psi$ ($^\circ$)       &    268.85             &             &   --                             \\
$\Omega($sr$)$          &    0.3920             &             &   --                             \\
$T_s/T_*$               &    0.9690             &             &   --                             \\
$\sum{\omega_i(O-C)_i^2}$ &  0.008574           &             &  0.020054                        \\
\hline\hline
\end{tabular}
\end{table*}

\section{Discussions and conclusion}

Combining the results $(M_1+M_2)sin^3i=1.330\pm0.012M_{\odot}$ and
$q_{sp}=0.455\pm0.011$ (Pribulla et al. 2006), absolute parameters
about each component were calculated to be,
$M_1=1.319\pm0.007M_{\odot}$, $M_2=0.600\pm0.01M_{\odot}$;
$R_1=1.28\pm0.19R_{\odot}$, $R_2=0.91\pm0.13R_{\odot}$;
$L_1=1.253L_{\odot}$, $L_2=1.997L_{\odot}$.

Based on all available photoelectric and CCD eclipse times, the
period changes of the contact binary star were discussed in the
previous section. The general O-C trend may reveal a long-term
period increase at a rate of
$dP/dt=+3.97\times{10^{-8}}$\,days/year. Meanwhile, a
small-amplitude period oscillation ($A_3=0.^{d}0026$) was discovered
superimposed on the period increase. If this period increase is due
to a conservative mass transfer from the less massive component to
the more massive one, then with the absolute parameters derived by
the present paper and by using the well-known equation,
\begin{equation}
\frac{\dot{P}}{P}=3\frac{\dot{M_2}}{M_2}(1-\frac{M_2}{M_1}),
\end{equation}
\noindent the mass transfer rate is estimated to be,
$dM_2/dt=3.82\times{10^{-8}}M_{\odot}/year$. The timescale of mass
transfer is $\tau\sim{M_2/\dot{M_2}}\sim1.58\times{10^{7}}$\,years
which is close to the thermal time scale of the secondary component.

\begin{table*}
\begin{minipage}{12cm}
\caption{The masses and orbital radii of the assumed third body in
TX Cnc.}
\begin{tabular}{lll}
      \hline\hline
Parameters & TX Cnc & Units\\
       \hline
$A_3$      & $0.0028(\pm0.0006)$   & d\\
$T_3$      & $26.6({\rm assumed})$ & yr\\
$e^{\prime}$ & $0({\rm assumed})$& ---\\
$a_{12}^{\prime}{\rm sin}i^{\prime}$ & $0.48(\pm0.10)$ & {\rm AU}\\
$f(m)$   & $1.56(\pm0.9)\times{10^{-4}}$ & $M_{\odot}$\\
$m_{3}(i^{\prime}=90^{\circ})$  & $0.086(\pm0.018)$ & $M_{\odot}$ \\
$m_{3}(i^{\prime}=70^{\circ})$  & $0.091(\pm0.019)$ &  $M_{\odot}$ \\
$m_{3}(i^{\prime}=50^{\circ})$  & $0.113(\pm0.066)$ &  $M_{\odot}$ \\
$m_{3}(i^{\prime}=30^{\circ})$  & $0.176(\pm0.011)$ &  $M_{\odot}$ \\
$m_{3}(i^{\prime}=10^{\circ})$  & $0.569(\pm0.040)$ &  $M_{\odot}$ \\
$a_{3}(i^{\prime}=90^{\circ})$  & $10.7(\pm3.5)$ &  AU \\
$a_{3}(i^{\prime}=70^{\circ})$  & $10.8(\pm3.3)$ &  AU \\
$a_{3}(i^{\prime}=50^{\circ})$  & $10.6(\pm2.8)$ &  AU \\
$a_{3}(i^{\prime}=30^{\circ})$  & $10.5(\pm2.0)$ &  AU \\
$a_{3}(i^{\prime}=10^{\circ})$  & $9.3(\pm1.0) $ &  AU \\
        \hline\hline
\end{tabular}
\end{minipage}
\end{table*}
As shown in Figures 2 and 3, both the primary and the secondary
times of light minimum follow the same general trend of O-C
variation indicating that the weak O-C oscillation can not be
explained as apsidal motion. The alternate period change of a close
binary containing at least one solar-type component can be
interpreted by the mechanism of magnetic activity (e.g., Applegate,
1992; Lanza et al. 1998). However, for contact binary stars, we do
not know whether this mechanism can work how does it might work,
because there is a common convective envelope. We think the period
oscillation may be caused by the light-time effect of a tertiary
component. As we can see from Figure 3, the data after E=27500 show
large scatters, therefore details on the information of orbital
eccentricity are unknown. In the previous section, by assuming a
circular orbit, a theoretical solution of the orbit for the assumed
tertiary star was calculated. By using this equation;
\begin{equation}
f(m)=\frac{4\pi^{2}} {{\it
G}T_3^{2}}\times(a_{12}^{\prime}\sin{i}^{\prime})^{3},
\end{equation}
where $a_{12}^{\prime}\sin{i}^{\prime}=A_3\times{c}$ (where c is the
speed of light), the mass function from the tertiary component is
computed. Then, with the following equation;
\begin{equation}
f(m)=\frac{(M_{3}\sin{i^{\prime}})^{3}} {(M_{1}+M_{2}+M_{3})^{2}}
\end{equation}
and taking the physical parameters given by us, the masses and the
orbital radii of the third companion are computed. The values of the
masses and the orbital radii of the third component stars for
several different orbital inclinations ($i^{\prime}$) are shown in
Table 5. As shown in this table, the assumed tertiary component is
invisible unless the orbital inclination $i^{\prime}$ is very small
($i^{\prime}<10^{\circ}$). If the tertiary companion is coplanar to
the eclipsing pair(i.e.,with the same inclination as the eclipsing
binary), its mass should be, $m_3=0.097M_{\odot}$, which is too
small to be detected. Actually, Pribulla et al. (2006) didn't
discover the third body, it maybe suspected or nondetected as they
said. More evidence is needed to show the existence of the third
body.

We estimated the orbital angular momentum and the spin angular
momentum of the system with the absolute parameters. As a crude
estimate, it assumes that the star is a rigid body rotator, using
the formula $J_{spin}=M_iR_i^2\omega$, where $\omega$ is the
self-rotation velocity dependent on the period. Meanwhile the
orbital momentum was calculated by;
\begin{equation}
J_{orb}=(GA)^{1/2}\frac{M_1M_2}{(M_1+M_2)^{1/2}},
\end{equation}
\noindent and the results are $J_{spin}=4.56\times10^{42}
kg{\cdot}m^2{\cdot}s^{-1}$, $J_{orb}=5.78\times10^{44}
kg{\cdot}m^2{\cdot}s^{-1}$. $J_{orb}$ is much larger than
$J_{spin}$. Hut (1980) pointed out a critical condition that if the
orbital angular momentum is less than three times the total spin
angular momentum, the system will become unstable and evolve into
single rapid-rotating stars. This is another approach which is
different from a period decrease system. But TX Cnc seems far from
that condition. Consider with that TX Cnc is a member of the young
cluster NGC2632(with an age of (3-5)$\times{10^{8}}$\,years), the
discussion above may indicate that TX Cnc just formed its contact
configuration.

Recent period studies by Qian (2001a,b; 2003a) have shown the
long-term period variation of contact binary stars may correlate
with the mass of the primary component ($M_{1}$) and with the mass
ratio of the system ($q$). Systems with higher $M_{1}$ and $q$
usually display an increasing period, the secular period increase of
TX Cnc is consistent with this conclusion. In order to interpret the
secular period changes of contact binary stars, an evolutionary
scenario was proposed by Qian (2001a, b, 2003a). According to this
scenario, the evolution of a contact binary may be the combination
of the thermal relaxation oscillation (TRO) and the variable angular
momentum loss (AML) via the change of depth of contact.  Systems
(e.g., V417 Aql, see Qian 2003b) with a secular decreasing period
are on the AML-controlled stage, while those (e.g., CE Leo, see Qian
2002) showing an increasing period are on the TRO-controlled stage.
The long-term period increase of TX Cnc may suggest that it is on
the TRO-controlled stage of this evolutionary scheme.

The high frequency of contact binaries in old open clusters have
been reported by several investigators (e.g., Kaluzny \& Shara 1987;
Kubiak et al. 1992; Kaluzny et al. 1993; Mazur, et al. 1995;
Rucinski 1998, Zhang et al. 2002). However, a survey made by Kaluzny
\& Shara (1988) of six open clusters with age no less than 4Gyr did
not find a single W UMa-type binary star. These properties are in
agreement with the formation of contact binary stars from detached
binaries by angular momentum loss via magnetic braking. This
mechanism was first proposed by Huang (1967) and was later
investigated by Van't Veer (1979); Rahunen (1981); Vilhu (1982);
Guinan \& Bradstreet (1988); Hilditch et al. (1988); Van't Veer \&
Maceroni (1989) and others. Praesepe (M44) is a young open cluster
with an age of (3-5)$\times{10^{8}}$\,years (e.g., Von Hoerner 1957;
Maeder 1971). The W-type contact binary star TX Cnc present in this
cluster makes it a very interesting system (Guinan \& Bradstreet
1988; Rucinski 1994). As discussed for AP Leo (Qian et al. 2007) and
AH Cnc ( Qian, et al. 2006), the tertiary component in TX Cnc, if it
really exists, may play an important role in the formation and
evolution of this binary star by removing a large amount of angular
momentum from the central system (Pribulla \& Rucinski, 2006). Thus
the system may have a short initial orbital period or a collision
path to fast evolution. Thus, the large disparity in age between TX
Cnc and almost all other contact binaries in other open clusters can
be interpreted.

\vskip 0.3in \noindent This work was partly supported by Yunnan
Natural Science Foundation (No.2005A0059M) \& the Chinese Natural
Science Foundation(10573032, 10573013, and 10433030). New
observations of TX Cnc were obtained with the 1.0-m telescope at
Yunnan Observatory. Thanks to the anonymous referee who given us
useful comments and cordial suggestions, which helped us to improve
the paper greatly.

\end{document}